\documentclass[12pt,aps,prb]{revtex4}
\usepackage{epsfig}
\usepackage{amsmath}

\normalsize

\def\b0{{\bf 0}}

\def\tG{\tilde G}

\def\Im{{\rm Im}}
\def\bra{\langle}
\def\ket{\rangle}
\def\up{\uparrow}
\def\down{\downarrow}

\def\alf{\alpha}
\def\eps{\epsilon}
\def\gam{\gamma}
\def\Gam{\Gamma}
\def\lam{\lambda}
\def\Lam{\Lambda}
\def\om{\omega}

\def\sg{\sigma}
\def\Sg{\Sigma}

\begin{document}


\title{\large 
  Renormalization group analysis of the one-dimensional \\
  extended Hubbard model with a single impurity}

 \author{S. Andergassen,$^1$ T. Enss,$^1$ V. Meden,$^2$ 
 W. Metzner,$^1$ \\
 U. Schollw\"ock,$^3$ and K. Sch\"onhammer$^2$ \\
 {\small\em $^1$Max-Planck-Institut f\"ur Festk\"orperforschung, 
 D-70569 Stuttgart, Germany} \\
 {\small\em $^2$Institut f\"ur Theoretische Physik, Universit\"at 
 G\"ottingen, D-37077 G\"ottingen, Germany} \\
 {\small\em $^3$Institut f\"ur Theoretische Physik C, RWTH Aachen, 
 D-52056 Aachen, Germany}}

\date{\small\today}


\begin{abstract}
We analyze the one-dimensional extended Hubbard model with a 
single static impurity by using a computational technique based 
on the functional renormalization group. 
This extends previous work for spinless fermions to 
spin-$\frac{1}{2}$ fermions.
The underlying approximations are devised for weak interactions 
and arbitrary impurity strengths, and have been checked by
comparing with density matrix renormalization group data.
We present results for the density of states, the density profile 
and the linear conductance.
Two-particle backscattering leads to striking effects,
which are not captured if the bulk system is approximated by
its low-energy fixed point, the Luttinger model.
In particular, the expected decrease of spectral weight near the 
impurity and of the conductance at low energy scales is often 
preceded by a pronounced increase, and the asymptotic power 
laws are modified by logarithmic corrections. \\
\mbox{PACS: 71.10.Pm, 73.21.Hb, 72.10.-d} \\
\end{abstract}

\maketitle



\section{Introduction}

One-dimensional metallic electron systems are always strongly 
affected by interactions. At low energy scales many observables
obey anomalous power laws, known as Luttinger-liquid behavior,
which is very different from conventional Fermi-liquid behavior
describing most higher dimensional metals.\cite{Gia,Voi}
For spin-rotation invariant systems all power-law exponents
can be expressed in terms of a single nonuniversal parameter
$K_{\rho}$.
For Luttinger liquids with repulsive interactions ($K_{\rho}<1$) 
already one static impurity has a strong effect at low energy
scales, even when the impurity potential is relatively 
weak.\cite{LP,Mat,AR,GS}
The asymptotic low-energy properties of Luttinger liquids with 
a single impurity have been investigated already in the 1990s.
\cite{KF,FN,YGM}
For electron systems (spin-$\frac{1}{2}$ fermions) with 
$K_{\rho}<1$ the essential properties can be summarized as follows.
The backscattering amplitude generated by a weak impurity is a 
relevant perturbation which grows as $\Lam^{(K_{\rho}-1)/2}$ 
for a decreasing energy scale $\Lambda$.
On the other hand, the tunneling amplitude through a weak link 
between two otherwise separate wires scales to zero as 
$\Lambda^{\alpha_B}$, with the boundary exponent 
$\alpha_B = (K_{\rho}^{-1}-1)/2$.
At low energy scales any impurity thus effectively ``cuts'' 
the system in two parts with open boundary conditions at 
the end points, and physical observables are controlled by the 
open chain fixed point.\cite{KF} In particular,
the local density of states near the impurity is suppressed as 
$\rho \sim |\omega|^{\alpha_B}$ for $|\omega| \to 0$, 
and the conductance vanishes as $G(T) \sim T^{2\alpha_B}$ at 
low temperatures.
We note that these power laws are strictly valid only in the
absence of two-particle backscattering. In general they are
modified by logarithmic corrections. 
The asymptotic behavior is universal in the sense that the exponents
depend only on the properties of the bulk system, via $K_{\rho}$,
while they do not depend on the impurity strength or shape, except
in special cases such as resonant scattering at double barriers, 
which require fine-tuning of parameters.

The progress in the fabrication of artificial low-dimensional 
structures stimulated advanced experimental verification of the 
theoretical predictions.\cite{KS} 
In an appropriate temperature and energy range Luttinger-liquid 
behavior can be expected in several systems with a predominantly 
one-dimensional character, such as organic conductors like the 
Bechgaard salts, artificial quantum wires in semiconductor
heterostructures or on surface substrates, carbon nanotubes, and 
fractional quantum Hall fluids for chiral Luttinger liquids.
For a correct interpretation of experimental data it would be
helpful to have theoretical input beyond asymptotic power laws,
which are valid only at sufficiently low energy scales.
It is not always clear whether the asymptotic Luttinger-liquid 
behavior is well developed before finite size effects and 
interactions with the three-dimensional environment become
important. 

Recently, a functional renormalization group (fRG) method has
been developed for a direct treatment of \emph{microscopic} models
of interacting fermions with impurities in one dimension 
\cite{MMSS,AEX,EMX}
which not only captures correctly the universal low-energy
behavior, but allows one to compute observables on all energy
scales, yielding thus also nonuniversal properties, and in 
particular an answer to the important question below which scale
the asymptotic power laws are actually valid.
The method has been applied to the spinless fermion model with 
nearest-neighbor interaction on a one-dimensional lattice, 
supplemented by various types of impurity potentials.
The most relevant observables such as the local density of
states,\cite{MMSS,AEX} the density profile,\cite{AEX} and the 
linear conductance \cite{EMX,MAX} were calculated.
The truncation of the fRG hierarchy of flow equations employed
in these works is valid only for sufficiently weak interactions.
However, a comparison with exact numerical results from the 
density matrix renormalization group (DMRG) \cite{dmrg} showed 
that the truncated flow equations are generally rather accurate 
also for sizable interaction parameters.
The fRG captures complex crossover phenomena at intermediate
scales, such as the temperature dependence of the conductance
through a resonant double barrier.\cite{EMX,MEX}
It can also be applied to other (than chain) geometries, such
as mesoscopic rings threaded by a magnetic flux \cite{MS} or 
Y junctions.\cite{BSMS}

In this work we extend the fRG method for interacting Fermi
systems with a single or few impurities to spin-$\frac{1}{2}$ 
fermions and apply it to the extended one-dimensional Hubbard 
model. For fermions with spin, vertex renormalization is 
crucial to take into account that two-particle backscattering 
of fermions with opposite spins at opposite Fermi points scales 
to zero in the low-energy limit. By contrast, for spinless 
fermions the effects of a single static impurity are captured
qualitatively already within the lowest order truncation of the 
fRG hierarchy of flow equations, where the renormalized vertex
is approximated by the bare interaction.
Two-particle backscattering scales to zero only logarithmically,
and thus gives rise to logarithmic corrections to the asymptotic 
power laws. 

The analysis of fixed-point models, which yields the ultimate
low-energy behavior, predicts a power-law decay of the local
density of states near an impurity or boundary of systems with
$K_{\rho} < 1$. However, a non-selfconsistent Hartree-Fock and 
a DMRG study of the density of states 
near the end of a finite Hubbard model chain with open boundary 
conditions revealed that the power-law suppression at the lowest
scales can be preceded by a pronounced increase of spectral
weight.\cite{SMX,MMX}
A similar crossover behavior can therefore also be expected
for the density of states near an impurity, at least for a
sufficiently strong one, as we indeed obtain in this work from
the fRG.
For the conductance, a renormalization group analysis of the 
g-ology model by Matveev et al.\cite{YGM} showed
that two-particle backscattering can lead to an increase as
a function of decreasing temperature before the asymptotic
suppression sets in.

The paper is organized as follows.
In Sec.\ II we introduce the microscopic model and derive the 
corresponding fRG flow equations.
In Sec.\ III we present results for spectral properties of 
single-particle excitations near an impurity or boundary, the 
density profile, and transport properties.
We conclude with a summary and an outlook in Sec.\ IV.

\section{Model and flow equations}
 
\subsection{Microscopic model}

As a microscopic model for the bulk electron system we choose the 
one-dimensional extended Hubbard model with a nearest-neighbor
hopping amplitude $t$, a local interaction $U$, and a 
nearest-neighbor interaction $U'$.
The bulk system is supplemented by a site or hopping impurity.
The total Hamiltonian is given by
\begin{equation}
 H = -t \sum_{j,\sg} \big( \,
 c^{\dag}_{j+1,\sg} c_{j\sg}^{\phantom{\dag}} + 
 c^{\dag}_{j\sg} c_{j+1,\sg}^{\phantom{\dag}} \, \big) \; + 
 U \sum_j n_{j\up} \, n_{j\down} + U' \sum_j n_j \, n_{j+1}
 \, + H_{\rm imp} \; ,
\end{equation}
where $c^{\dag}_{j\sg}$ and $c_{j\sg}^{\phantom{\dag}}$ are 
creation and annihilation operators for fermions with spin 
projection $\sg$ on site $j$, while
$n_{j\sg} = c^{\dag}_{j\sg} \, c_{j\sg}^{\phantom{\dag}} \,$,
and $n_j = n_{j\up} + n_{j\down}$ is the local density operator. 
For the (nonextended) Hubbard model the nearest-neighbor 
interaction vanishes.
A local site impurity on site $j_0$ is modeled by 
$H_{\rm imp} = V \, n_{j_0}$, 
and a hopping impurity by the nonlocal potential 
$H_{\rm imp} = (t-t') \sum_{\sg} \big( \,
 c^{\dag}_{j_0+1,\sg} c_{j_0,\sg}^{\phantom{\dag}} 
+ c^{\dag}_{j_0,\sg} \, c_{j_0+1,\sg}^{\phantom{\dag}} \, \big)$, 
such that the hopping amplitude $t$ is replaced by $t'$ on the bond 
linking the sites $j_0$ and $j_0+1$. 
In the following we will set the bulk hopping amplitude $t$ equal
to one, that is all energies are expressed in units of $t$.

In the absence of impurities, the Hubbard model can be solved exactly 
using the Bethe-ansatz,\cite{LW} while the extended Hubbard model 
is not integrable. 
The Hubbard model is a Luttinger liquid for arbitrary repulsive 
interactions at all particle densities except half-filling, where 
the system becomes a Mott insulator.\cite{Gia,Voi} 
The phase diagram of the extended Hubbard model is more complex.
Away from half-filling, it is a Luttinger liquid at least for 
sufficiently weak repulsive interactions.\cite{Voi}
For the Hubbard model the Luttinger-liquid parameter $K_{\rho}$
can be computed exactly from the Bethe ansatz solution.\cite{Sch}

For the calculation of transport properties a finite interacting 
chain (with sites $1,\dots,L$) is coupled to noninteracting leads 
at both ends. 
The influence of the leads on the interacting chain can be taken 
into account by incorporating a dynamical boundary potential
\begin{equation} \label{vlead}
 V_j^{\rm lead}(i\omega_n) = 
 \frac{i\omega_n+\mu_0}{2} \left( 1 - 
 \sqrt{1 - \frac{4}{(i\omega_n+\mu_0)^2}} \, \right) 
 \left( \delta_{1,j} + \delta_{L,j} \right) \; ,
\end{equation}
in the bare propagator $G_0$ of the interacting chain.\cite{EMX} 
The parameter $\mu_0$ is the chemical potential, which is related 
to the density $n$ in the leads by $\mu_0 = -2 \cos k_F$ with 
$k_F = n\pi/2$.
Uncontrolled conductance drops due to scattering at the contacts
between leads and the interacting part of the chain can be avoided
by switching off the interaction potential smoothly near the 
contacts. In addition, interaction induced bulk shifts of the 
density have to be compensated by a suitable bulk potential.
\cite{EMX}

\subsection{Flow equations}

We now extend the fRG scheme derived and used
previously for spinless Fermi systems with impurities
\cite{MMSS,AEX,EMX} to electrons, that is spin-$\frac{1}{2}$ 
fermions. We make use of equations and procedures described
already in detail in the articles Ref.~\onlinecite{AEX} and 
Ref.~\onlinecite{EMX}, without repeating the derivations here.

We use the one-particle irreducible (1PI) version of the fRG.
\cite{Wet,Mor,SH}
The starting point is an exact hierarchy of differential flow 
equations for the 1PI vertex functions, 
which is obtained by introducing an infrared cutoff $\Lam$ in 
the free propagator and differentiating the effective action 
with respect to $\Lam$. 
Since translation invariance is spoiled by the impurity, we use
a Matsubara frequency cutoff instead of a cutoff on momenta.
The cutoff is sharp at $T=0 \,$ \cite{AEX} and smooth for 
$T>0 \,$.\cite{EMX}
The hierarchy is truncated by neglecting the contribution of
the three-particle vertex to the flow of the two-particle vertex. 
The coupled system of flow equations for the two-particle vertex
$\Gam^{\Lam}$ and the self-energy $\Sg^{\Lam}$ is then closed.
The contribution of the three-particle vertex to $\Gam^{\Lam}$
is small as long as $\Gam^{\Lam}$ is sufficiently small.

We neglect the influence of the impurity on the flow of the
two-particle vertex, such that $\Gam^{\Lam}$ remains translation
invariant. While this is sufficient for capturing the effects
of isolated impurities in otherwise pure systems, it is known
that impurity contributions to vertex renormalization become
important in macroscopically disordered systems.\cite{Gia}
We also neglect the feedback of the bulk self-energy into the
flow of $\Gam^{\Lam}$, which yields only a very small 
correction at weak coupling.
The two-particle vertex is parametrized approximately by a
renormalized static short-range interaction \cite{AEX}
in order to reduce the number of variables in the flow,
which would be unmanageably large otherwise.
This approximation is exact at the beginning of the flow and
fully captures the nonirrelevant parts of the vertex in the
low-energy limit.
The self-energy generated by the simplified vertex is then
static (frequency independent) and its spatial dependence 
can be treated fully, that is without resorting to another
simplified parametrization.
Transport properties are computed by coupling the interacting
model to noninteracting leads as described in Ref.~\onlinecite{EMX}.
The conductance is obtained directly from the one-particle
Green function, since current vertex corrections vanish in our
approximation for $\Gam^{\Lam}$.

We now describe the parametrization of the spatial (or momentum)
dependences of the two-particle vertex $\Gam^{\Lam}$ for 
spin-$\frac{1}{2}$ fermions, employing a natural extension
of our previous parametrization for the spinless case.\cite{AEX}
We consider only spin-rotation invariant lattice systems with 
local and nearest-neighbor interactions. This includes the
extended Hubbard model. 

For a spin-rotation invariant system the spin structure of the
two-particle vertex can be decomposed into a singlet and a triplet 
part: 
\begin{equation}
  \label{eq:st}
 \Gam^{\Lam} = 
 \Gam_s^{\Lam} \, S_{\sg'_1,\sg'_2;\sg_1,\sg_2} +
 \Gam_t^{\Lam} \, T_{\sg'_1,\sg'_2;\sg_1,\sg_2}
\end{equation}
with
\begin{eqnarray}
 S_{\sg'_1,\sg'_2;\sg_1,\sg_2} &=& \frac{1}{2} \,
 \left( \delta_{\sg_1\sg'_1} \delta_{\sg_2\sg'_2} -
        \delta_{\sg_1\sg'_2} \delta_{\sg_2\sg'_1} \right)
 \nonumber \\
 T_{\sg'_1,\sg'_2;\sg_1,\sg_2} &=& \frac{1}{2} \,
 \left( \delta_{\sg_1\sg'_1} \delta_{\sg_2\sg'_2} +
        \delta_{\sg_1\sg'_2} \delta_{\sg_2\sg'_1} \right) \; .
\end{eqnarray}
Since the total vertex is antisymmetric in the incoming and outgoing
particles, the singlet part $\Gam_s^{\Lam}$ has to be symmetric and
the triplet part $\Gam_t^{\Lam}$ antisymmetric.

Proceeding in analogy to the case of spinless fermions,\cite{AEX} 
we first list momentum components of the vertex with all momenta 
at $\pm k_F$.
For the triplet vertex the antisymmetry allows only one such 
component
\begin{equation}
\label{eq:gt}
 g^{\Lam}_t = \Gam^{\Lam}_{t|\, k_F,-k_F;k_F,-k_F} \; .
\end{equation}
For the singlet vertex there are several distinct components at
$\pm k_F$.
Since we will neglect the influence of the impurity on the
vertex renormalization, the renormalized vertex remains translation
invariant. Hence the momentum components are restricted by momentum 
conservation: $k'_1 + k'_2 = k_1 + k_2$, modulo integer multiples 
of $2\pi$.
The remaining independent (not related by obvious symmetries) 
components are
\begin{eqnarray}
 g^{\Lam}_{s2} &=& \Gam^{\Lam}_{s|\, k_F,-k_F;k_F,-k_F} \; ,
 \nonumber \\[2mm]
 g^{\Lam}_{s4} &=& \Gam^{\Lam}_{s|\, k_F,k_F;k_F,k_F} \; ,
\end{eqnarray}
and in the case of half-filling, for which $k_F = \pi/2$, also
\begin{equation}
 g^{\Lam}_{s3} = \Gam^{\Lam}_{s|\, \pi/2,\pi/2;-\pi/2,-\pi/2} \; .
\end{equation}
The labels $2,3,4$ are chosen in analogy to the conventional 
g-ology notation for one-dimensional Fermi systems.\cite{Sol}
In order to parametrize the vertex in a uniform way in all cases,
we will include the umklapp component $g^{\Lam}_{s3}$ not only
at half-filling, but at any density. The effect on the other components 
is negligible for the range of interactions and fillings considered.

Extending our treatment of the spinless case,\cite{AEX} we now 
parametrize the vertex by renormalized local and nearest-neighbor 
interactions in real space. 
For the triplet part, there is no local component, and only one 
nearest-neighbor component compatible with the antisymmetry, 
namely
\begin{equation}
 {U'_t}^{\Lam} = \Gam^{\Lam}_{t|\, j,j+1;j,j+1} \; ,
\end{equation}
which has the same form as the nearest-neighbor interaction in the 
spinless case. Note that $\Gam^{\Lam}_{t|\, j,j+1;j,j+1}$ does not
depend on $j$, and is equal to $\Gam^{\Lam}_{t|\, j,j-1;j,j-1}$.
For the symmetric singlet part, there is one local component
\begin{equation}
 U_s^{\Lam} = \Gam^{\Lam}_{s|\, j,j;j,j} \; ,
\end{equation}
and three different components involving nearest neighbors:
\begin{eqnarray}
 {U'_s}^{\Lam} &=& \Gam^{\Lam}_{s|\, j,j+1;j,j+1} \nonumber \\[2mm]
 P^{\Lam}_s    &=& \Gam^{\Lam}_{s|\, j+1,j+1;j,j} \nonumber \\[2mm]
 W^{\Lam}_s    &=& \Gam^{\Lam}_{s|\, j+1,j;j,j} \; .
\end{eqnarray}
For the Hubbard model, the bare vertex is purely local and the
initial condition for the vertex is given by $U_s^{\Lam_0} = 2U$,
while all the other components vanish. For the extended Hubbard model, 
${U'_s}^{\Lam_0}={U'_t}^{\Lam_0}=U'$ are nonzero.

The triplet vertex is parametrized by only one renormalized real 
space coupling, which leads to a momentum representation of the form
\begin{equation}
  \label{eq:paramt}
  \Gam^{\Lam}_{t|\, k'_1,k'_2;k_1,k_2} =
  2 {U'_t}^{\Lam} \, [ \cos(k'_1 - k_1) - \cos(k'_2 - k_1) ] \,
  \delta^{(2\pi)}_{k_1+k_2,k'_1+k'_2} \; ,
\end{equation}
where the Kronecker delta implements momentum conservation 
(modulo $2\pi$).
The flowing coupling ${U'_t}^{\Lam}$ is thus linked in a one-to-one 
correspondence to the Fermi momentum coupling $g^{\Lam}_t$ by
\begin{equation}
  \label{eq:paramgt}
  g^{\Lam}_t = 2 {U'_t}^{\Lam} \, [ 1 - \cos(2k_F) ] \; ,
\end{equation}
as in the spinless case.\cite{AEX}
In the singlet channel we have found four real space couplings,
that is one more than necessary to match the three singlet couplings 
in momentum space, $g^{\Lam}_{s2}$, $g^{\Lam}_{s3}$, $g^{\Lam}_{s4}$. 
We choose to discard the interaction $W^{\Lam}_s$,
because it does not appear in the bare Hubbard model,
where it is generated only at third order in $U$,
while the pair hopping $P^{\Lam}_s$ appears already in second order
perturbation theory.
Fourier transforming the remaining interactions yields the singlet 
vertex in k-space
\begin{equation}
  \label{eq:params}
 \Gam^{\Lam}_{s|\, k'_1,k'_2;k_1,k_2} =
 \left[ U_s^{\Lam} + 
 2 {U'_s}^{\Lam} \, [ \cos(k'_1 - k_1) + \cos(k'_2 - k_1) ] +
 2 P^{\Lam}_s \, \cos(k_1 + k_2) \right] \,
 \delta^{(2\pi)}_{k_1+k_2,k'_1+k'_2}
\end{equation}
from which we obtain a linear relation between the momentum space
couplings $g^{\Lam}_{s2}$, $g^{\Lam}_{s3}$, $g^{\Lam}_{s4}$ and the 
renormalized interaction parameters $U_s^{\Lam}$, ${U'_s}^{\Lam}$, 
$P^{\Lam}_s \,$:
\begin{eqnarray}
  \label{eq:paramgs}
  g^{\Lam}_{s2} &=& U_s^{\Lam} + 
  2 {U'_s}^{\Lam} \, [1 + \cos(2k_F)] + 2 P^{\Lam}_s 
  \nonumber \\[2mm]
  g^{\Lam}_{s3} &=& U_s^{\Lam} - 4 {U'_s}^{\Lam} - 2 P^{\Lam}_s
  \nonumber \\[2mm]
  g^{\Lam}_{s4} &=& U_s^{\Lam} + 4 {U'_s}^{\Lam} +
  2 P^{\Lam}_s \, \cos(2k_F) \; .
\end{eqnarray} 
The determinant of this linear system is positive for all $k_F$,
except for $k_F = 0$ and $\pi$.
Hence the equations can be inverted for all densities except the 
trivial cases of an empty or completely filled band.

We can now set up the flow equations for the four independent 
couplings ${U'_t}^{\Lam}$, $U_s^{\Lam}$, ${U'_s}^{\Lam}$, and
$P^{\Lam}_s$ which parametrize the vertex. 
Consider the case $T=0$ first.
Inserting the spin structure (\ref{eq:st}) into the general flow 
equation for the two-particle vertex, Eq.~(18) in 
Ref.~\onlinecite{AEX},
and using the momentum representation for a translation invariant 
vertex, the flow equation for the singlet and triplet vertices
$\Gam^{\Lam}_a$, $a = s,t$, can be written as
\begin{equation}
  \label{eq:flow}
  \frac{\partial}{\partial\Lam} \, 
  \Gam^{\Lam}_{a|\, k'_1,k'_2;k_1,k_2} =
  - \frac{1}{2\pi} \sum_{\om = \pm\Lam} \, \sum_{b,b' = s,t} 
  \int \frac{dp}{2\pi} \, (\, {\rm PP + PH + PH'} \,)
\end{equation}
with the particle-particle and particle-hole contributions
\begin{eqnarray}
 {\rm PP} &=& C^{\rm PP}_{a,bb'} \, 
 G^0_p(i\om) \, G^0_{k_1+k_2-p}(-i\om) \,
 \Gam^{\Lam}_{b|\, k'_1,k'_2;p,k_1+k_2-p} \, 
 \Gam^{\Lam}_{b'|\, p,k_1+k_2-p;k_1,k_2} 
 \nonumber \\[2mm]
 {\rm PH} &=& C^{\rm PH}_{a,bb'} \, 
 G^0_p(i\om) \, G^0_{p+k_1-k'_1}(i\om) \,
 \Gam^{\Lam}_{b|\, k'_1,p+k_1-k'_1;k_1,p} \, 
 \Gam^{\Lam}_{b'|\, p,k'_2;p+k_1-k'_1,k_2} 
 \nonumber \\[2mm]
 {\rm PH'} &=& C^{\rm PH'}_{a,bb'} \, 
 G^0_p(i\om) \, G^0_{p+k_1-k'_2}(i\om) \,
 \Gam^{\Lam}_{b|\, k'_2,p+k_1-k'_2;k_1,p} \, 
 \Gam^{\Lam}_{b'|\, p,k'_1;p+k_1-k'_2,k_2} \; .
\end{eqnarray}
The coefficients $C_{a,bb'}$ are obtained from the spin sums as
\begin{equation}
 \begin{array}{ll}
 C^{\rm PP}_{s,ss} = 1 \, , \; &
 C^{\rm PP}_{s,st} = C^{\rm PP}_{s,ts} = C^{\rm PP}_{s,tt} = 0
 \\[2mm]
 C^{\rm PP}_{t,tt} = 1 \, , \; &
 C^{\rm PP}_{t,ss} = C^{\rm PP}_{t,st} = C^{\rm PP}_{t,ts} = 0
 \\[2mm]
 C^{\rm PH}_{s,ss} = -1/4 \, , \; &
 C^{\rm PH}_{s,st} = C^{\rm PH}_{s,ts} = C^{\rm PH}_{s,tt} = 3/4
 \\[2mm]
 C^{\rm PH}_{t,tt} = 5/4 \, , \; &
 C^{\rm PH}_{t,ss} = C^{\rm PH}_{t,st} = C^{\rm PH}_{t,ts} = 1/4
 \\[2mm]
 C^{\rm PH'}_{s,bb'} = - \, C^{\rm PH}_{s,bb'} \, , \; &
 C^{\rm PH'}_{t,bb'} = C^{\rm PH}_{t,bb'} \; .
 \end{array}
\end{equation}
Note that we have neglected the self-energy feedback in the flow
of $\Gam^{\Lam}$, such that only bare propagators $G_0$ enter. 
On the right hand side of the flow equation we insert the 
parametrization (\ref{eq:paramt}) for $\Gam^{\Lam}_t$ and 
(\ref{eq:params}) for $\Gam^{\Lam}_s$.
The flow of the triplet vertex $\Gam^{\Lam}_{t|\, k'_1,k'_2;k_1,k_2}$
is evaluated only for $(k'_1,k'_2,k_1,k_2) = (k_F,-k_F,k_F,-k_F)$ as 
in (\ref{eq:gt}), which yields the flow of $g^{\Lam}_t$, while 
the flow of the singlet vertex $\Gam^{\Lam}_{s|\, k'_1,k'_2;k_1,k_2}$
is computed for the three choices of $k'_1,k'_2,k_1,k_2$ which yield
the flow of $g^{\Lam}_{s2}$, $g^{\Lam}_{s3}$, $g^{\Lam}_{s4}$.
Using the linear equations (\ref{eq:paramgt}) and (\ref{eq:paramgs}) 
to replace the couplings $g^{\Lam}$ by the renormalized real space 
interactions on the left hand side of the flow equations, we obtain 
a complete set of flow equations for the four renormalized 
interactions ${U'_t}^{\Lam}$, $U_s^{\Lam}$, ${U'_s}^{\Lam}$, and 
$P^{\Lam}_s$ of the form
\begin{equation} \label{eq:flowu}
 \partial_{\Lam} U^{\Lam}_{\alf} = 
 \sum_{\om = \pm\Lam} \sum_{\alf', \alf''}
 h_{\alf' \alf''}^{\alf}(\om) \, 
 U^{\Lam}_{\alf'} \, U^{\Lam}_{\alf''} \; ,
\end{equation}
where $\alf = 1,2,3,4$ labels the four different interactions. 
The functions $h_{\alf'\alf''}^{\alf}(\om)$ can be computed 
analytically by carrying out the momentum integrals in (\ref{eq:flow})
via the residue theorem; the flow equations can then be solved 
numerically very easily. 
The expressions become too lengthy to be reported here; for details 
we refer the interested reader to the thesis by one of the authors. 
\cite{And}
For finite systems the momentum integral should be replaced by a 
discrete momentum sum; however, this leads only to negligible 
corrections for the physical observables presented in Sec.~III.

After computing the flow of the real space interactions, one can
also calculate the flow of the momentum space couplings $g^{\Lam}$
by using the linear relation between the two.
In the low-energy limit (small $\Lam$) one recovers the one-loop
flow of the g-ology model, the general effective low-energy model for 
one-dimensional fermions.\cite{Sol}
In addition, our vertex renormalization captures also all nonuniversal
second order contributions to the vertex at $\pm k_F$ from higher energy
scales.

\begin{figure}[ht!]
\center{\includegraphics[width=10.cm]{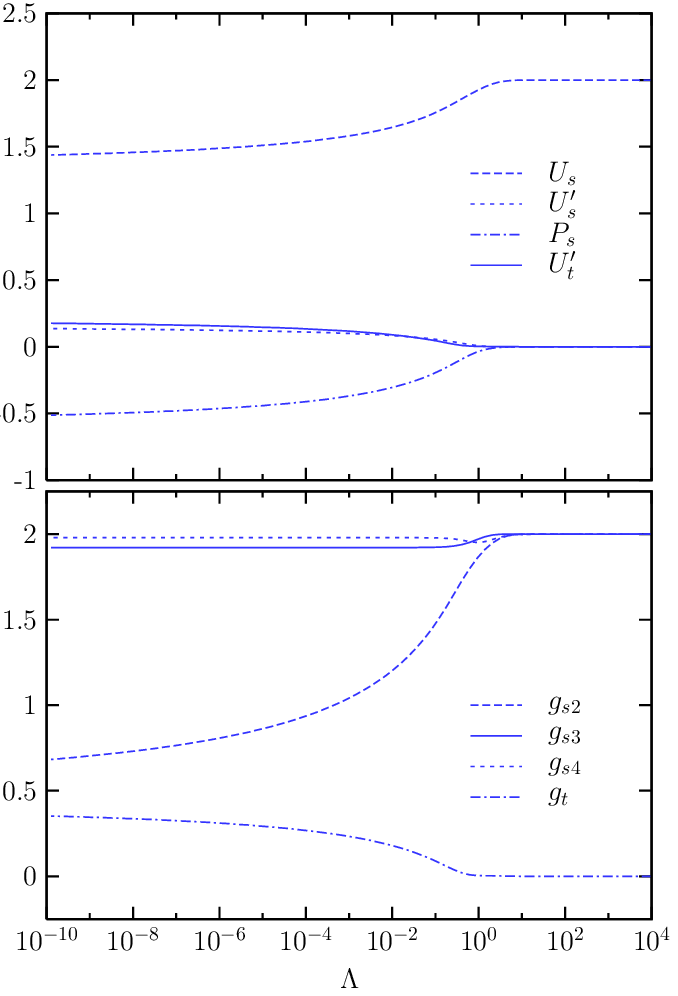}}
\vskip 5mm 
\caption{\label{fig:gy} (Color online)
 Vertex flow for the Hubbard model at 
 quarter-filling ($n=1/2$) and $U = 1$; 
  \emph{upper panel}: 
  flow of the renormalized real space interactions, 
  \emph{lower panel}: 
  flow of the momentum space couplings.}
\end{figure}

In Fig.~\ref{fig:gy} we show results for the renormalized real space 
interactions together with the corresponding momentum space couplings,
as obtained by integrating the flow equations for the Hubbard model at 
quarter-filling and $T=0$. 
Note that the couplings converge to finite fixed-point values in 
the limit $\Lam \to 0$, but the convergence is very slow, 
except for the momentum space couplings $g^{\Lam}_{s3}$ and 
$g^{\Lam}_{s4}$.
This can be traced back to the familiar behavior of the so-called
backscattering coupling 
$g^{\Lam}_{1\perp} = \frac{1}{2} \, (g^{\Lam}_{s2} - g^{\Lam}_t)$,
that is the amplitude for the exchange of two particles with opposite 
spin at opposite Fermi points.
Backscattering is known to vanish logarithmically in the low-energy 
limit for spin-rotation invariant spin-$\frac{1}{2}$ Luttinger 
liquids.\cite{Voi}
We emphasize that this logarithmic behavior is not promoted to a 
power law by higher order terms beyond our approximation.
By contrast, the linear combination of couplings which determines 
the Luttinger-liquid parameter $K_{\rho}$ converges very quickly 
to a finite fixed-point value (see below).

Due to the above parametrization of the vertex by real space 
interactions which do not extend beyond nearest neighbors on the 
lattice, the self-energy generated by the flow equations is 
frequency independent and tridiagonal in real space.
Inserting the spin and real space structure of $\Gamma^{\Lambda}$ 
into the general flow equation for the self-energy, Eq.~(16)
in Ref.~\onlinecite{AEX}, one obtains
\begin{eqnarray} \label{eq:flows}
 \frac{\partial}{\partial\Lam} \, \Sg^{\Lam}_{j,j} &=&
 - \frac{1}{4\pi} \sum_{\om = \pm\Lam} \Big[ \,
 U_s^{\Lam} \, \tG^{\Lam}_{j,j}(i\om) +
 ({U'_s}^{\Lam} + 3 {U'_t}^{\Lam}) 
 \sum_{r = \pm 1} \, \tG^{\Lam}_{j+r,j+r}(i\om) \, \Big]
 \nonumber \\[2mm]
 \frac{\partial}{\partial\Lam} \, \Sg^{\Lam}_{j,j \pm 1} &=&
 - \frac{1}{4\pi} \sum_{\om = \pm\Lam} \left[ \,
 ({U'_s}^{\Lam} - 3 {U'_t}^{\Lam}) \, \tG^{\Lam}_{j,j \pm 1}(i\om)
 + P^{\Lam}_s \, \tG^{\Lam}_{j \pm 1,j}(i\om) \, \right] \; ,
\end{eqnarray}
where $\tG^{\Lam} = (G_0^{-1} - \Sg^{\Lam})^{-1}$.
These equations can be solved very efficiently,\cite{Ens}
so that very large systems with up to $10^7$ sites can be
treated.

At $T>0$ the Matsubara frequencies are discrete, and a sharp frequency
cutoff therefore leads to discontinuities in the flow.  
To avoid ambiguities and numerical problems associated with these
discontinuities one may choose a smooth frequency cutoff as in 
Ref.~\onlinecite{EMX}.
Alternatively, one can rewrite the Matsubara sum as a frequency
integral with a suitable weight function, as described in detail in
the Appendix. The latter procedure leads to a particularly simple
and numerically convenient extension of the flow equations to finite 
temperatures.
In the flow equations (\ref{eq:flowu}) and (\ref{eq:flows}) one has 
to replace $\om = \pm\Lam$ by $\om = \pm\om_n^{\Lam}$, where 
$\om_n^{\Lam}$ is the Matsubara frequency which is closest to $\Lam$. 
The function $h_{\alf' \alf''}^{\alf}(\om)$ remains the same.
At $T>0$ the flow equations can be solved for systems with up to $10^5$ 
sites without extensive numerical effort.

When using a sharp cutoff with $\tG^\Lam$ defined above it makes no
difference whether the bare impurity potential is put into $G_0^{-1}$
or $\Sg^\Lam$; we choose to include it in the initial condition of 
$\Sg^{\Lam}$ at $\Lam = \infty$.
Due to the slow decay of $G$ at large frequencies, the integration 
of the flow equation for $\Sg$ from $\Lam = \infty$ to 
$\Lam = \Lam_0$ yields a contribution which remains finite even in 
the limit $\Lam_0 \to \infty$.\cite{AEX} 
For the extended Hubbard model this contribution is given by
$\Sg^{\Lam_0}_{j,j} = U/2 + 2U'$ for $j = 2, \dots, L \!-\! 1$
and $\Sg^{\Lam_0}_{1,1} = \Sg^{\Lam_0}_{L,L} = U/2 + U'$. 
The numerical integration of the flow is started at a sufficiently 
large $\Lam_0$ with $\Sg^{\Lam_0}$ as initial condition.

\subsection{Calculation of ${\boldsymbol K_{\bf\rho}}$}

The Luttinger-liquid parameter $K_{\rho}$ can be computed from the 
fixed-point couplings as obtained from the fRG. 
A relation between the fixed-point couplings and $K_{\rho}$ can be
established via the exact solution of the fixed-point Hamiltonian 
of Luttinger liquids, the Luttinger model.
Since the above simplified flow equations yield not only the correct 
low-energy asymptotics to second order in the renormalized interaction,
but contain also all {\em nonuniversal}\/ second order corrections 
at $\pm k_F$ at higher energy scales, the resulting $K_{\rho}$
is obtained correctly to second order in the interaction. 

The Luttinger-liquid parameter $K_{\rho}$ is given by 
\begin{equation}
 K_{\rho} = \sqrt{\frac{1 + (g_{\rho 4} - g_{\rho 2})/(\pi v_F)}
 {1 + (g_{\rho 4} + g_{\rho 2})/(\pi v_F)}} \; .
\end{equation}
The coupling constants $g_{\rho 2}$ and $g_{\rho 4}$ parametrize 
forward scattering interactions in the charge channel (which is 
spin symmetrized) between opposite and equal Fermi points, 
respectively. They are related to the bare singlet and triplet 
vertices of the Luttinger model by
\begin{eqnarray}
 g_{\rho 2} &=& \frac{1}{4} \, 
 \big(\, \gam_{s|\, k_F,-k_F;k_F,-k_F} + 
 3 \gam_{t|\, k_F,-k_F;k_F,-k_F} \, \big)
 \nonumber \\[2mm]
 g_{\rho 4} &=& \frac{1}{4} \, \gam_{s|\, k_F,k_F;k_F,k_F} \; .
\end{eqnarray}
These bare vertices are identical to the \emph{dynamical} forward 
scattering limits of the full vertex $\Gam$.
On the other hand, the vertex $\Gam^{\Lam}$ obtained from the fRG
with a frequency cutoff yields the \emph{static} forward scattering
limit for $\Lam \to 0$.\cite{AEX} 
For the Luttinger model, the static forward scattering limit of
the vertex can be computed from the effective interactions 
$D_{\rho 2}(q,i\nu)$ and $D_{\rho 4}(q,i\nu)$, which are defined 
as the sum over all particle-hole chains with the bare interactions 
$g_{\rho 2}$ and $g_{\rho 4}$.\cite{Sol}
The summation becomes a simple geometric series if one introduces 
symmetric and antisymmetric combinations
$g_{\rho\pm} = g_{\rho 4} \pm g_{\rho 2}$ and 
$D_{\rho\pm}(q,i\nu) = D_{\rho 4}(q,i\nu) \pm D_{\rho 2}(q,i\nu)$.
The static limit of the effective interaction $D_{\rho\pm}(q,i\nu)$ 
yields the relation
\begin{equation}
  \label{eq:gpm}
 g^*_{\rho\pm} = \frac{g_{\rho\pm}}{1 - g_{\rho\pm}/(\pi v_F)}
\end{equation}
between the Luttinger-model couplings $g_{\rho\pm}$ and the 
fixed-point couplings
\begin{equation}
 g^*_{\rho\pm} = \frac{1}{4} \, \big[ \,
 g^*_{s4} \pm \big( \, g^*_{s2} + 3 g^*_t \, \big)  \big]
\end{equation}
from the fRG with frequency cutoff. Inverting (\ref{eq:gpm}) one 
obtains
\begin{equation}
 K_{\rho} = \sqrt{\frac{1 - g^*_{\rho +}/(\pi v_F)}
 {1 - g^*_{\rho -}/(\pi v_F)}} \; .
\end{equation}
The Fermi velocity $v_F$ can be computed from the self-energy for 
the translation invariant pure system as in the spinless case,
\cite{AEX} using the momentum representation of the flow 
equations (\ref{eq:flows}). 

The results for $K_{\rho}$ from the above procedure are correct to 
second order in the bare interaction for the Hubbard model and also 
for the extended Hubbard model.
While the flowing couplings $g^{\Lam}_{s2}$ and $g^{\Lam}_t$
converge only logarithmically to their fixed-point values for
$\Lam \to 0$, the linear combination $g^{\Lam}_{s2} + 3 g^{\Lam}_t$
which enters $K_{\rho}$ converges much faster.

\begin{figure}[ht!]
\center{\includegraphics[width=10.cm]{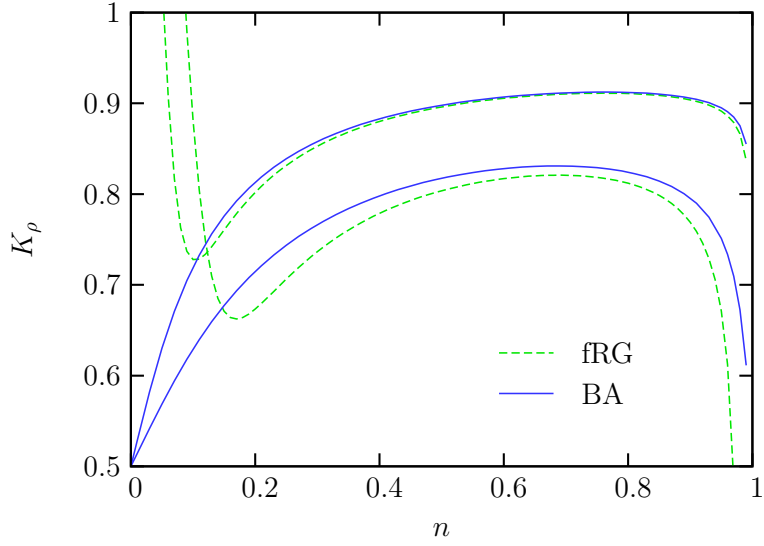}}
\vskip 5mm 
\caption{\label{fig:tk_n} (Color online)
 Luttinger-liquid parameter $K_{\rho}$ 
 for the Hubbard model as a function of electron density.
 Results from the fRG are compared to exact results from the
 Bethe ansatz.
 The upper curves are for $U = 1$ and the lower ones for $U = 2$.}
\end{figure}

In Fig.~\ref{fig:tk_n} we show results for $K_{\rho}$ for the 
Hubbard model as obtained from the fRG and, for comparison, from 
the exact Bethe ansatz solution.\cite{Sch}
The truncated fRG yields accurate results at weak coupling except 
for low densities and close to half-filling. 
In the latter case this failure is expected since umklapp 
scattering interactions renormalize toward strong coupling,
even if the bare coupling is weak. At low densities already 
the bare dimensionless coupling $U/v_F$ is large for fixed finite
$U$, simply because $v_F$ is proportional to $n$ for small $n$, 
such that neglected higher order terms become important. 
Note, for comparison, that for spinless fermions with a fixed 
nearest-neighbor interaction the bare dimensionless coupling 
at the Fermi level vanishes in the low-density limit.

\begin{figure}[ht!]
\center{\includegraphics[width=10.cm]{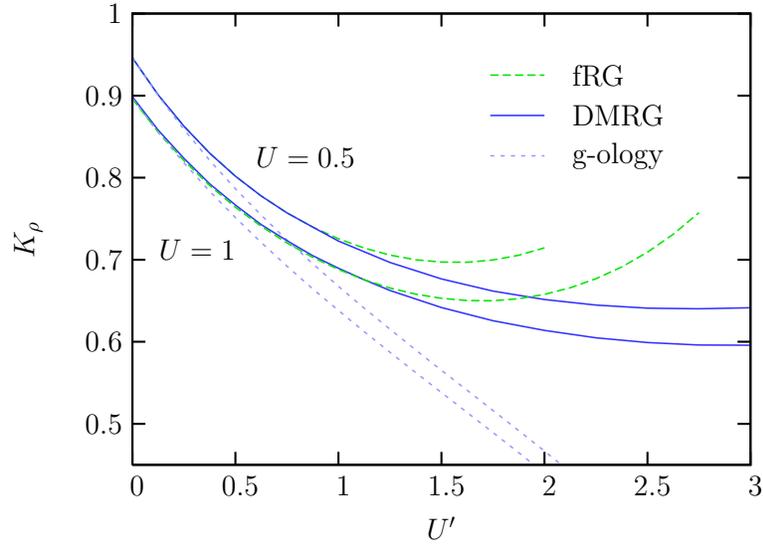}}
\vskip 5mm 
\caption{\label{fig:tkr} (Color online)
 Luttinger-liquid parameter $K_{\rho}$ for 
 the quarter-filled extended Hubbard model as a function of $U'$ 
 for $U = 0.5$ and $U = 1$. Results from the fRG are compared to
 DMRG data and to results from a one-loop g-ology calculation.}
\end{figure}

For the extended Hubbard model Fig.~\ref{fig:tkr} shows a comparison 
of fRG results for $K_{\rho}$ to DMRG data.\cite{EGN}
The fRG results are exact to second order in the interaction 
and are thus very accurate for weak $U$ and $U'$. 
Results from a standard one-loop g-ology calculation \cite{Sol}
deviate quite strongly already for $U' > 0.5$. 
In the g-ology approach interaction processes are classified 
into backward scattering $(g_{1 \perp})$, forward scattering 
involving electrons from opposite Fermi points $(g_{2 \perp})$, 
from the same Fermi points $(g_{4\perp})$, and umklapp scattering 
$(g_{3 \perp})$. 
All further momentum dependences of the vertex are discarded.
This is justified by the irrelevance of these momentum dependences
in the low-energy limit, but leads to deviations from the exact 
flow at finite scales, and therefore to less accurate results for 
the fixed-point couplings.

\begin{figure}[ht!]
\center{\includegraphics[width=10.cm]{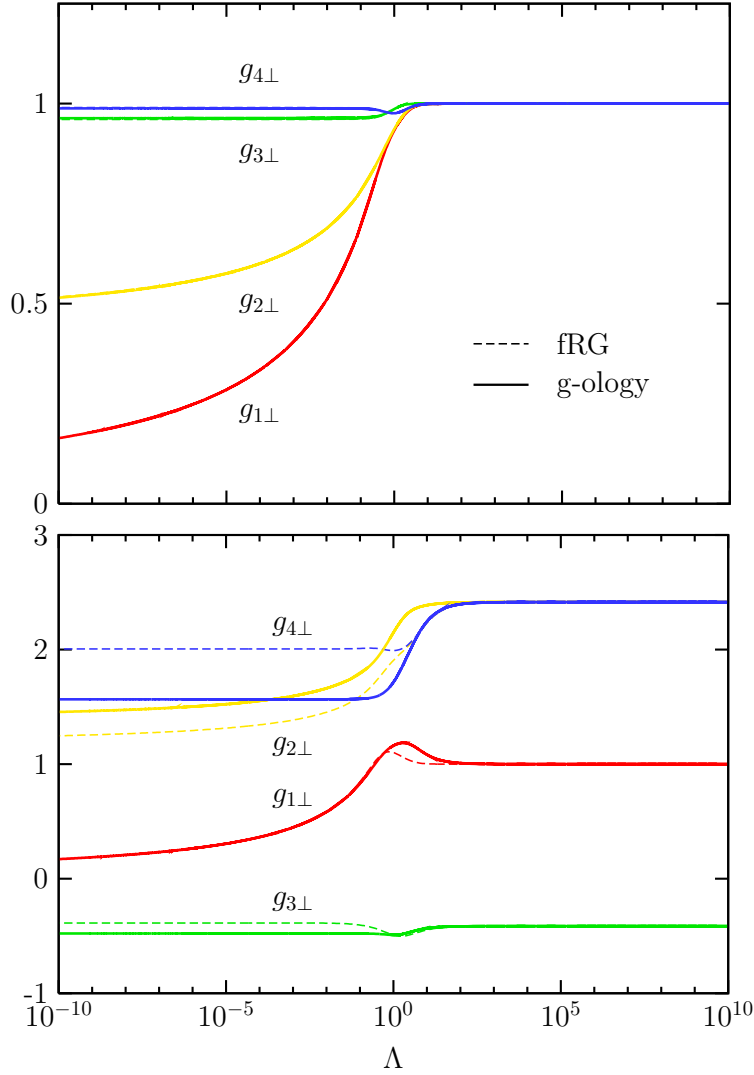}}
\vskip 5mm 
\caption{\label{fig:gort} (Color online)
 Flow of the vertex on the Fermi points
 (in g-ology notation) at quarter-filling and $U = 1$;
  \emph{upper panel}: Hubbard model,  
  \emph{lower panel}: extended Hubbard model with $U' = U/\sqrt{2}$.
 The fRG flow is compared to the one-loop g-ology flow;
 note that in the upper panel fRG and g-ology results almost
 coincide.}
\end{figure}

The flow of $g_{i\perp}^{\Lambda}$, $i=1,...,4$, is plotted in 
Fig.~\ref{fig:gort}, in the upper panel for the quarter-filled
Hubbard model with bare interaction $U=1$, and for the extended 
Hubbard model with $U'=U/\sqrt{2}$ in the lower.
The fRG result is compared to the result from a one-loop g-ology
calculation.
The backscattering coupling $g_{1 \perp}$ vanishes logarithmically
in both cases, as expected for the Luttinger-liquid fixed point.\cite{Voi}
For the pure Hubbard model the good agreement with g-ology results 
stems from the purely local interaction in real space, since in
that case pronounced momentum dependences of the vertex develop
only in the low-energy regime where the g-ology parametrization
is a good approximation.
By contrast, for the extended Hubbard model momentum dependences
of the vertex which are not captured by the g-ology classification
(except for small $\Lam$) are obviously more important.
A generalization of the g-ology parametrization of the vertex to 
higher dimensions, which amounts to neglecting the momentum 
dependence normal to the Fermi surface, is frequently used in 
one-loop fRG calculations in two dimensions.\cite{2dHubb}
The above comparison indicates that this parametrization works
well for the pure Hubbard model, but could be improved for models
with nonlocal interactions. The parametrization of the vertex
by an effective short-range interaction used here could be easily
extended to higher dimensions, where it will probably yield more
accurate results, too.
The relevance of an improved parametrization of the vertex
beyond the conventional g-ology classification has also been
demonstrated in a recent fRG analysis of the 
phase-diagram of the half-filled extended Hubbard model.\cite{TTC}

\section{Results for observables}

In this section we present and discuss explicit results for the
spectral properties of single-particle excitations, the density 
profile, and the conductance for the Hubbard and extended Hubbard 
model with a single impurity, as obtained from the solution of 
the fRG flow equations.
We also analyze excitations and density oscillations near a
boundary, which corresponds to an infinite site impurity or a
vanishing weak link.
A comparison with DMRG results is made for the spectral weight 
at the Fermi level and for the density profile.
For details on the computation of the relevant observables from the 
solution of the flow equations we refer to Ref.~\onlinecite{AEX,EMX}.

\subsection{Single-particle excitations}

Integrating the flow equation for the self-energy $\Sg^{\Lam}$
down to $\Lam = 0$ yields the physical self-energy $\Sg$ and the 
single-particle propagator $G = (G_0^{-1} - \Sg)^{-1}$.
From the Green function $G$ the properties of single-particle
excitations can be extracted.
We focus on the local spectral function given by
\begin{equation}
 \rho_j(\om) = - \frac{1}{\pi} \, 
 \Im \, G_{jj}(\om + i0^+) \; .
\end{equation}
For a finite system this function is a finite sum of $\delta$-peaks
of weight $w_{\lam j}$, where $\lam$ labels the eigenvalues of
the effective single-particle Hamiltonian defined by $G$.
Dividing the spectral weight $w_{\lam j}$ by the level spacing yields
the local density of states $D_j(\om)$.
Even-odd effects due to finite-size details in the spectral weight 
are averaged out by averaging over neighboring eigenvalues.

For $\om \to 0$ the spectral weights and the local density of
states near a boundary or impurity are ultimately suppressed 
according to a power law with the boundary exponent
\begin{equation} \label{eq:ab}
 \alpha_B = \frac{1}{2K_{\rho}}+\frac{1}{2K_{\sigma}} - 1
\end{equation}
with $K_{\sigma} = 1$ for spin-rotation invariant systems.
\cite{Gia}
However, due to the slow logarithmic decrease of the two-particle
backscattering amplitude, the fixed-point value of $K_{\sigma}$
is reached only logarithmically from above. 
Hence, we can expect that the asymptotic value of $\alpha_B$ is 
usually reached only very slowly from below.

\begin{figure}[ht!]
\center{\includegraphics[width=10.cm]{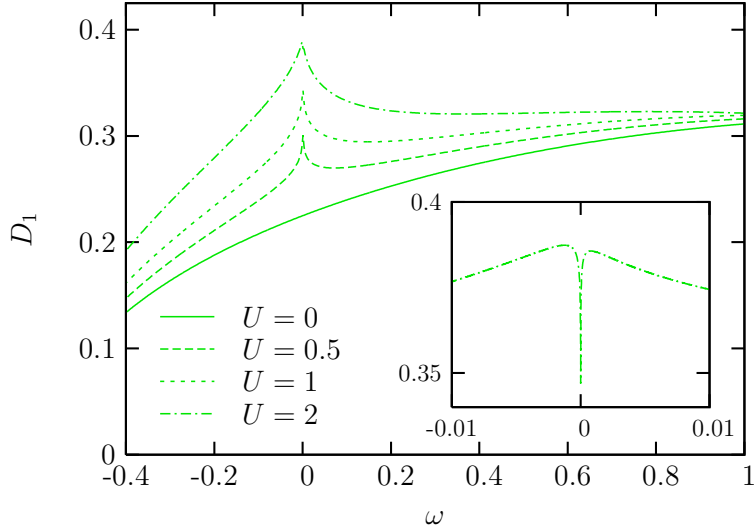}}
\vskip 5mm
\caption{\label{fig:dosa} (Color online)
 Local density of states at the boundary 
 of a Hubbard chain of length $L = 4096$ at quarter-filling
 and various interaction strengths $U$; 
 the inset shows results for $U=2$ and $L = 10^6$ at very low
 $\om$.}
\end{figure}
\begin{figure}[ht!]
\center{\includegraphics[width=10.cm]{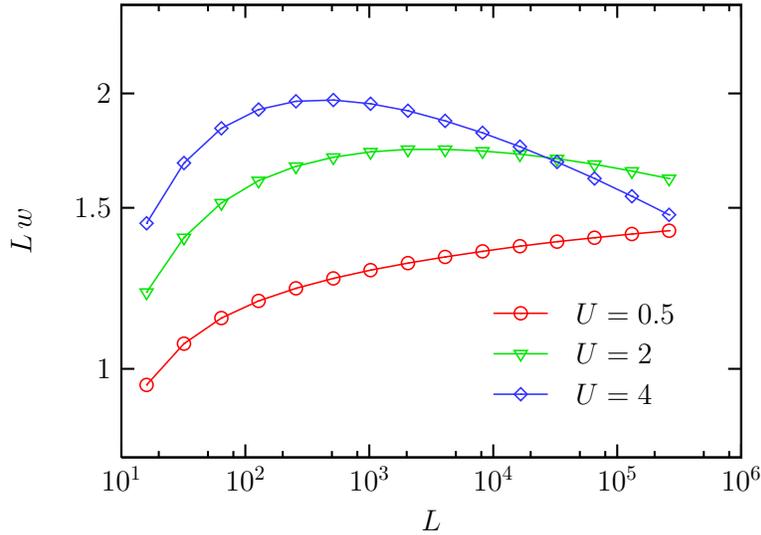}}
\vskip 5mm
\caption{\label{fig:twspin} (Color online)
 Spectral weight at the Fermi level 
 at the boundary of a quarter-filled Hubbard chain as a function 
 of system size $L$, for various different interaction strengths.}
\end{figure}

The local density of states at the boundary of a quarter-filled
Hubbard chain, computed by the fRG, is shown in Fig.~\ref{fig:dosa} 
for various values of the local interaction $U$. 
Contrary to the expected asymptotic power-law suppression, the 
spectral weight near the chemical potential is strongly 
\emph{enhanced}. 
The predicted suppression occurs only at very small energies for
sufficiently large systems.
In the main panel of Fig.~\ref{fig:dosa} the crossover to the 
asymptotic behavior cannot be observed, as the finite size cutoff 
$\sim \pi v_F/L$ is too large.
Results for a larger system with $L = 10^6$ sites at $U=2$ in 
the inset show the crossover to the asymptotic suppression, 
albeit only at very small energies.
The dependence of the boundary spectral weight at the Fermi level 
on the system size $L$ is plotted in Fig.~\ref{fig:twspin}. 
The $L$-dependence of the spectral weight at zero energy is expected 
to display the same asymptotic power-law behavior for large $L$ 
as the $\omega$-dependence discussed above.
Instead of decreasing with increasing $L$, the spectral weight
increases even for rather large systems for small and moderate 
values of $U$. For $U>2$ the crossover to a suppression is visible
in Fig.~\ref{fig:twspin}.
For $U=0.5$ only an increase is obtained up to the largest systems
studied.
The crossover scale depends sensitively on the interaction strength 
$U$; for small $U$ it is exponentially large in $v_F/U$.


The above behavior of the spectral weight and density of states
near a boundary of the Hubbard chain, that is a pronounced increase
preceding the asymptotic power-law suppression, is captured
qualitatively even by the Hartree-Fock approximation.\cite{SMX,MMX} 
This is at first sight surprising, as the Hartree-Fock theory does
not capture any Luttinger-liquid features in the bulk of a 
translation invariant system.
The initial increase of $D_j(\om)$ near a boundary is actually
obtained already within perturbation theory at first order in
the interaction,\cite{MMX}
\begin{equation}
 D_j(\om) = D_j^0(\om) \, \left[ \, 1 +
 \frac{\tilde V(0) - z \tilde V(2k_F)}{2\pi v_F} \, 
 \ln|\om/\eps_F| + {\cal O}(\tilde V^2) \, \right] \; ,
\end{equation}
where $D_j^0(\om)$ is the noninteracting density of states,
$\tilde V(q)$ the Fourier transform of the real space interaction,
and $z$ the number of spin components.
For spinless fermions ($z=1$) with repulsive interactions the
coefficient in front of the logarithm is always positive such
that the first order term leads to a suppression of $D_j(\om)$.
For the Hubbard model, one has $z=2$ and
$\tilde V(0) - 2 \tilde V(2k_F) = - U$ is negative for repulsive
$U$. Hence, at least for weak $U$ the density of states increases
for decreasing $\om$ until terms beyond first order become 
important.
For the extended Hubbard model,
$\tilde V(0) - 2 \tilde V(2k_F) = 2U'[1 - 2\cos(2k_F)] - U$,
which can be positive or negative for $U,U'>0$, 
depending on the density and the relative strength of the two
interaction parameters. 
At quarter-filling $\tilde V(0) - 2 \tilde V(2k_F)$ is negative 
and therefore leads to an enhanced density of states for $U' < U/2$.

Using g-ology notation, one can write
$\tilde V(0) - 2 \tilde V(2k_F) = g_{2\perp} - 2g_{1\perp}$, 
which reveals that substantial two-particle backscattering 
($g_{1\perp} > g_{2\perp}/2$) is necessary to obtain an enhancement 
of $D_j(\om)$ for repulsive interactions.
Backscattering vanishes at the Luttinger-liquid fixed point,
but only very slowly.
In case of a negative $\tilde V(0) - 2 \tilde V(2k_F)$ the
crossover to a suppression of $D_j(\om)$ is due to higher order
terms, which are expected to become important when the first 
order correction is of order one, that is for energies below
the scale
\begin{equation}
 \om_c = \eps_F \exp \left( 
 \frac{2\pi v_F}{\tilde V(0) - 2 \tilde V(2k_F)} \right) \; ,
\end{equation}
corresponding to a system size $L_c = \pi v_F/\om_c$.
The scale $\om_c$ is exponentially small for weak interactions.
A more accurate analytical estimate of the crossover scale
from enhancement to suppression has been derived for the
Hubbard model within Hartree-Fock approximation in 
Ref.~\onlinecite{MMX}. 
In a renormalization group treatment $\om_c$ is somewhat enhanced 
by the downward renormalization of backscattering.

\begin{figure}[ht!]
\center{\includegraphics[width=10.cm]{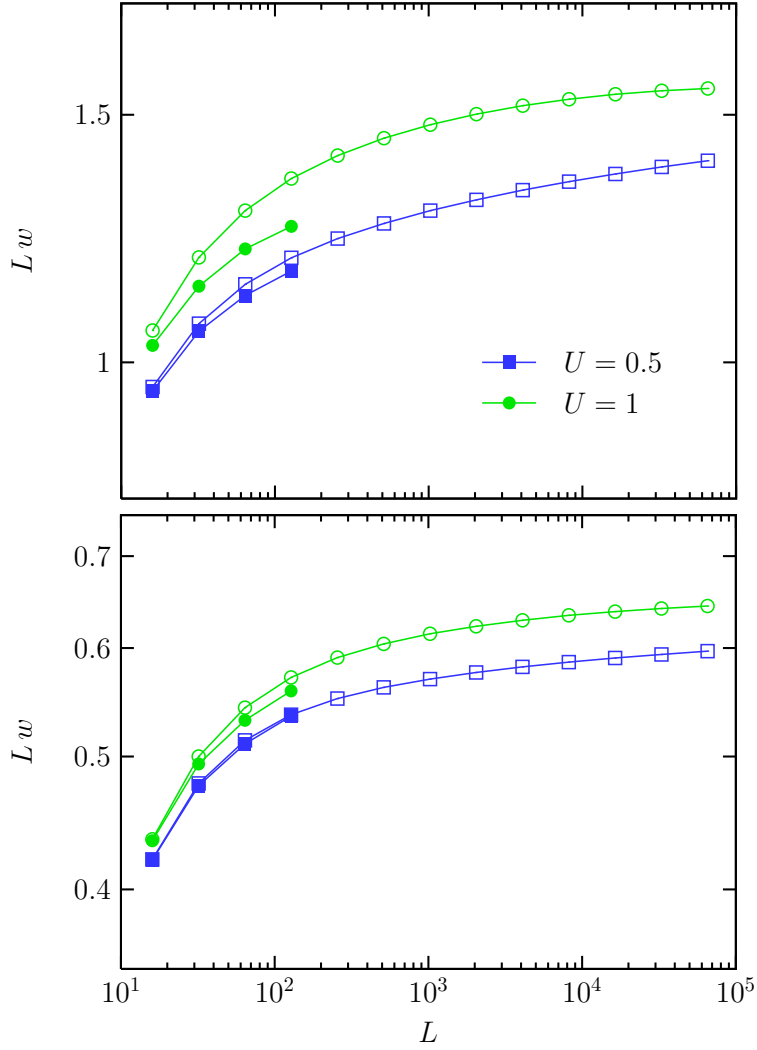}}
\vskip 5mm
\caption{\label{fig:whm} (Color online)
 Spectral weight at the Fermi level near a 
 boundary (\emph{upper panel}) and a hopping impurity $t'=0.5$ 
 (\emph{lower panel}) as a function of system size $L$ for the 
 Hubbard model at quarter-filling and different interaction 
 strengths $U$; results from the fRG (open symbols) are compared
 to DMRG data (filled symbols).}
\end{figure}

A comparison of fRG results with DMRG data \cite{dmrgspec} 
for the spectral weight at the Fermi level is shown in 
Fig.~\ref{fig:whm}, for a boundary site in the upper panel, and
near a hopping impurity of strength $t'=0.5$ in the lower. 
The agreement improves at weaker coupling, as expected, and is 
generally better for the impurity case, compared to the boundary
case.
The larger errors in the boundary case are probably due to our
approximate translation invariant parametrization of the 
two-particle vertex.
Boundaries and to a minor extent impurities spoil the translation 
invariance of the two-particle vertex.
Although the deviations from translation invariance of the
vertex become irrelevant in the low-energy or long-distance
limit, and therefore do not affect the asymptotic behavior,
they are nevertheless present at intermediate scales.
This feedback of impurities into the vertex increases of course
with the impurity strength and is thus particularly important 
near a boundary.
The scale for the crossover from enhancement to suppression of
spectral weight discussed above depends sensitively on effective
interactions at intermediate scales and can therefore be shifted
considerably even by relatively small errors in that regime.

\begin{figure}[ht!]
\center{\includegraphics[width=10.cm]{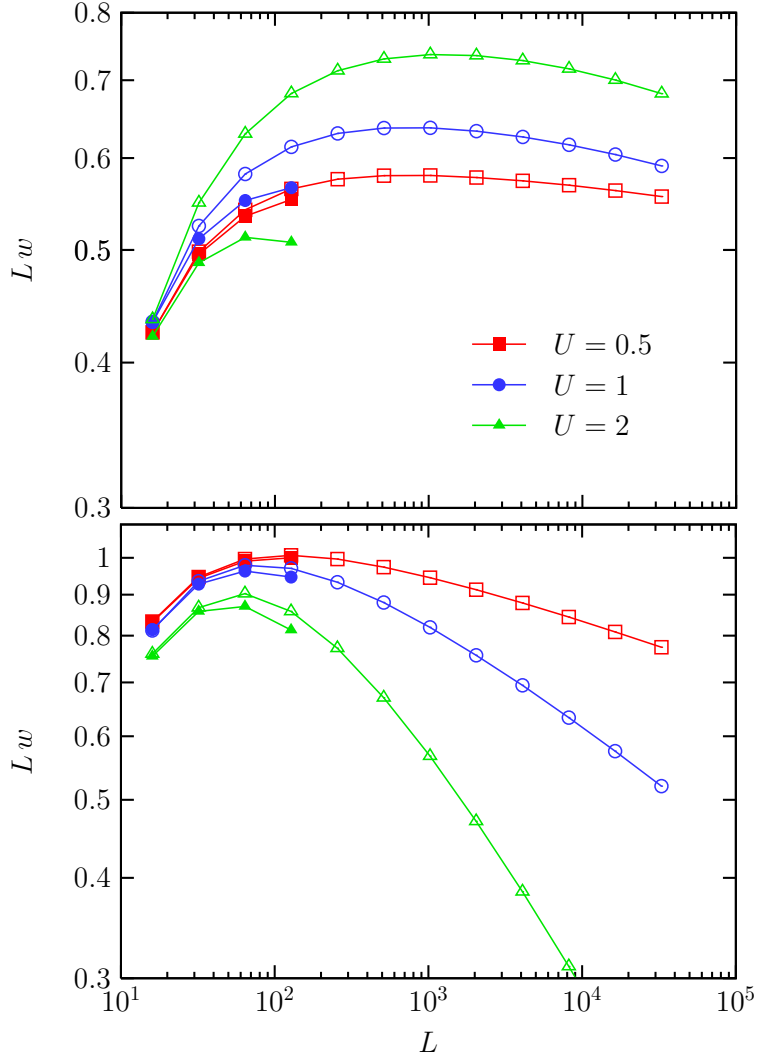}}
\vskip 5mm
\caption{\label{fig:w_gology} (Color online)
 Spectral weight at the Fermi level
 near a hopping impurity $t'=0.5$ as a function of system size $L$
 for the extended Hubbard model with $U' = U/\sqrt{2}$, 
 for various choices of $U$;
 \emph{upper panel}: $n = 1/2$ (leading to sizable backscattering),
 \emph{lower panel}: $n = 3/4$ (leading to small backscattering);
 results from the fRG (open symbols) are compared to DMRG data 
 (filled symbols).}
\end{figure}
\begin{figure}[ht!]
\center{\includegraphics[width=10.cm]{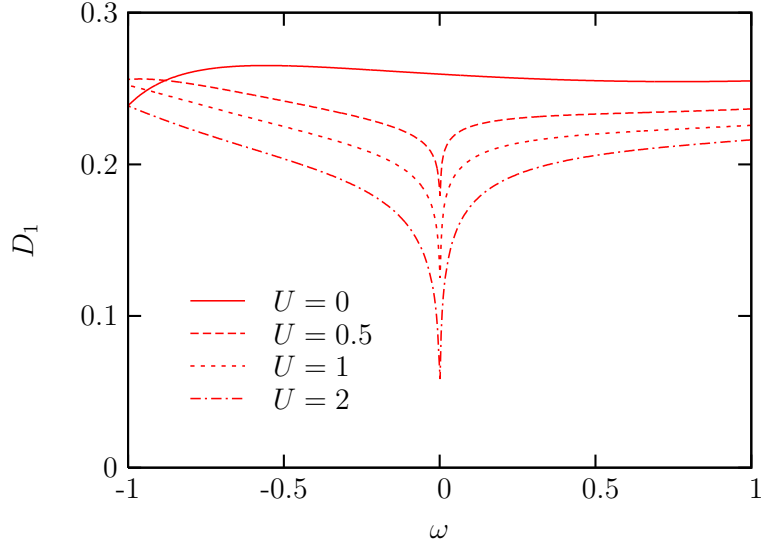}}
\vskip 5mm
\caption{\label{fig:dosc} (Color online)
 Local density of states near a hopping
 impurity $t' = 0.5$ in an extended Hubbard model with density 
 $n=3/4$ and interaction $U' = U/\sqrt{2}$ 
 (leading to a small backscattering interaction) 
 for various choices of $U$.
 The size of the chain is $L = 4096$.}
\end{figure}

With the additional nearest-neighbor interaction in the extended 
Hubbard model it is possible to tune parameters such that the
two-particle backscattering amplitude becomes negligible.
In that case the asymptotic power-law suppression of spectral
weight should be free from logarithmic corrections and accessible 
already for smaller systems and at higher energy scales.
The bare backscattering interaction in the extended Hubbard model 
is given by $g_{1\perp} = U + 2U'\cos(2k_F)$ and therefore vanishes
for $U' = - U/[2\cos(2k_F)]$, which is repulsive for $U > 0$ if 
$n > 1/2$.
In a one-loop calculation a slightly different value of $U'$ has to 
be chosen to obtain a negligible renormalized $g_{1\perp}^{\Lam}$
for small finite $\Lam$, since the flow generates backscattering
terms at intermediate scales even if the bare $g_{1\perp}$ vanishes.
In Fig.~\ref{fig:w_gology} we show fRG and DMRG results \cite{dmrgspec} 
for the spectral weight of the extended Hubbard model at the Fermi level
near a hopping impurity. In the upper panel a generic case with
sizable backscattering is shown, while the parameters leading 
to the curves in the lower panel have been chosen such that the
two-particle backscattering amplitude is negligible at low
energy.
Only in the latter case a pronounced suppression of spectral
weight is reached already for intermediate system size, similar
to the behavior obtained previously for spinless fermions with 
nearest-neighbor interaction.\cite{MMSS,AEX}
This is also reflected in the energy dependence of the local 
density of states near the impurity. 
For parameters leading to negligible two-particle backscattering 
as in Fig.~\ref{fig:dosc} the suppression of the density of 
states sets in already at relatively high energies and is not 
preceded by any interaction-induced increase.
Note also that the fRG results are much more accurate for small 
backscattering, as can be seen by comparing the agreement with
DMRG data in the upper and lower panel of Fig.~\ref{fig:w_gology}
especially for larger $U$.
This indicates that the influence of the impurity on the vertex 
flow, which we have neglected, is more important in the presence 
of a sizable backscattering interaction.

\begin{figure}[ht!]
\center{\includegraphics[width=10.cm]{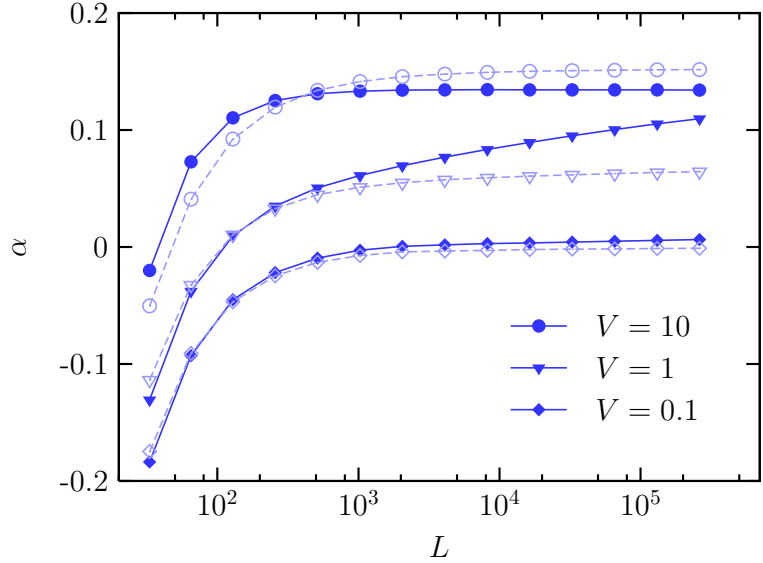}}
\vskip 5mm
\caption{\label{fig:timp1} (Color online)
 Logarithmic derivative of the spectral 
 weight at the Fermi level on the site next to a site impurity of 
 strength $V$ in the center of the chain as a function of system 
 size $L$, for the extended Hubbard model with interaction
 parameters $U = 1$, $U' = 0.65$, and density $n = 3/4$; 
 here the filled symbols are fRG, the open symbols Hartree-Fock 
 results.}
\end{figure}

In the case of a negligible backscattering amplitude, the 
spectral weight at the Fermi level approaches a power law without
logarithmic corrections for accessible system sizes if the
impurity is sufficiently strong.
The power law is seen most clearly by plotting the effective
exponent $\alpha(L)$, that is the negative logarithmic derivative 
of the spectral weight with respect to the system size.
Fig.~\ref{fig:timp1} shows $\alpha(L)$ on the site next to a site 
impurity of strength $V$ for the extended Hubbard model with 
$U = 1$, $U' = 0.65$, and $n = 3/4$. The backscattering 
amplitude is very small for these parameters.
The fRG results approach the expected universal $V$-independent
power law for large $L$, but only very slowly for small $V$. 
For a weak bare impurity potential $V$, the crossover to a strong 
effective impurity occurs only on a large length scale of order
$V^{2/(K_{\rho} - 1)} \,$.\cite{KF}
For $V=0.1$ this scale is obviously well above the largest system
size reached in Fig.~\ref{fig:timp1}.
The Hartree-Fock approximation also yields power laws for large
$L$, but the exponents depend on the impurity parameters.
This failure of Hartree-Fock theory was already observed earlier 
for spinless fermions.\cite{MMSS}

The effective exponent obtained from the fRG calculation agrees
with the exact boundary exponent to linear order in the bare
interaction, but not to quadratic order. 
To improve this, the frequency dependence of the two-particle 
vertex, which generates a frequency dependence of the self-energy, 
has to be taken into account. This is also necessary to describe
inelastic processes and to capture the anomalous dimension of the 
bulk system.
These effects could be included in an improved scheme by inserting 
the second order vertex into the flow equation for the self-energy 
without neglecting its frequency dependence.

\subsection{Density profile}

Boundaries and impurities induce a density profile with long-range 
Friedel oscillations, which are expected to decay as a power law
with exponent $(K_{\rho} + K_{\sigma})/2$ at long distances,
where $K_{\sg} = 1$ for spin-rotation invariant systems.\cite{EG}
For weak impurities linear response theory predicts a decay as 
$|j-j_0|^{1 - K_{\rho} - K_{\sg}}$ at intermediate distances. 

The density profile has to be computed from an additional flow 
equation, since the local density is a composite operator whose
renormalization is not well described by the propagator obtained
from the truncated flow of $\Sg^{\Lam}$.
The flow equation for $n_j^{\Lam}$ can be derived by computing the 
shift of the grand canonical potential $\Omega^{\Lambda}$ generated 
by a small field $\phi_j$ coupled to the local density. Its 
general structure at $T=0$ is described in Ref.~\onlinecite{AEX}. 
From Eqs.~(38) and (40) in that article one can easily obtain the
concrete flow equation for the case of the extended Hubbard model.

\begin{figure}[ht!]
\center{\includegraphics[width=10.cm]{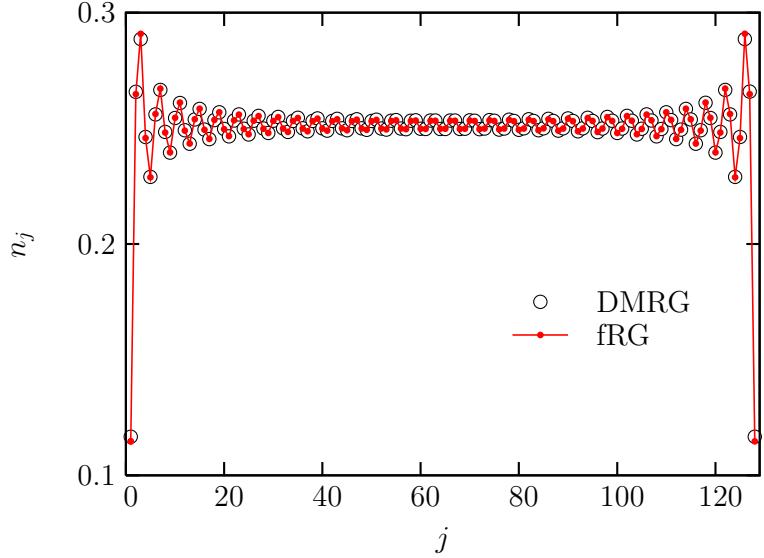}}
\vskip 5mm
\caption{\label{fig:tdensity128_1} (Color online)
 Density profile $n_j$ for the 
 Hubbard model with $128$ sites and 
 interaction strength $U = 1$ at quarter-filling;
 fRG results are compared to DMRG data.}
\end{figure}

As an additional benchmark for the fRG technique, we compare in 
Fig.~\ref{fig:tdensity128_1} fRG and DMRG results for the density 
profile $n_j$ for a quarter-filled Hubbard chain with $L=128$ 
lattice sites and open boundaries. 
Friedel oscillations emerge from both boundaries and interfere in 
the center of the chain. 
The fRG results have been shifted by a small constant amount to 
allow for a better comparison of the oscillations. Note that the
mean value of $n_j$ in the tails of the oscillations deviates from
the average density by a finite size correction of order $1/L$, 
which is related to the asymmetry of the oscillations near the 
boundaries.

The long-distance behavior of the density oscillations as obtained
within the fRG scheme has been analyzed in detail for spinless
fermions in Ref.~\onlinecite{AEX}.
For fermions with spin, asymptotic power laws can be identified 
only for special parameters leading to negligible two-particle 
backscattering. In general, the asymptotic behavior of Friedel 
oscillations is realized only at very long distances, and the
power laws are modified by logarithmic corrections.

\subsection{Conductance}

For the computation of the conductance a finite interacting chain 
is connected to two semi-infinite noninteracting leads, with a 
smooth decay of the interaction at the contacts.
The presence of leads modifies the propagator in the interacting 
region only via the boundary potential $V^{\rm lead}$, 
Eq.~(\ref{vlead}).
In linear response the conductance is given by \cite{EMX,Oguri}
\begin{eqnarray}
  \label{conductf}
  G(T) = - \frac{2e^2}{h} \int_{-2-\mu_0}^{2-\mu_0}
  |t(\varepsilon,T)|^2  \; f'(\varepsilon) \; d \varepsilon
\end{eqnarray} 
with $|t(\varepsilon,T)|^2 = [4- (\mu_0 + \varepsilon)^2] \, 
|G_{1,L}(\varepsilon,T)|^2$, and $f$ the Fermi function.
The factor two is due to the spin degeneracy. 
Within our approximation scheme, $\Sigma^{\Lambda}$ has no imaginary 
part, which implies that there are no vertex corrections, such that
the conductance is fully determined by boundary matrix element of the
single-particle propagator $G_{1,L}(\varepsilon,T)$.
\cite{Oguri}

For a system of spinless fermions with a single impurity it was 
already shown that the conductance obtained from the truncated fRG 
obeys the expected power laws, in particular $G(T) \propto 
T^{2\alf_B}$ at low $T$, and one-parameter scaling behavior.
\cite{EMX,MAX}
The corresponding scaling function agrees remarkably well with
an exact result for $K_{\rho} = 1/2$, although the interaction
required to obtain such a small $K_{\rho}$ is quite strong.
The more complex temperature dependence of the conductance in
the case of a double barrier at or near a resonance is also
fully captured by the fRG.\cite{EMX,MEX}

\begin{figure}[ht!]
\center{\includegraphics[width=10.cm]{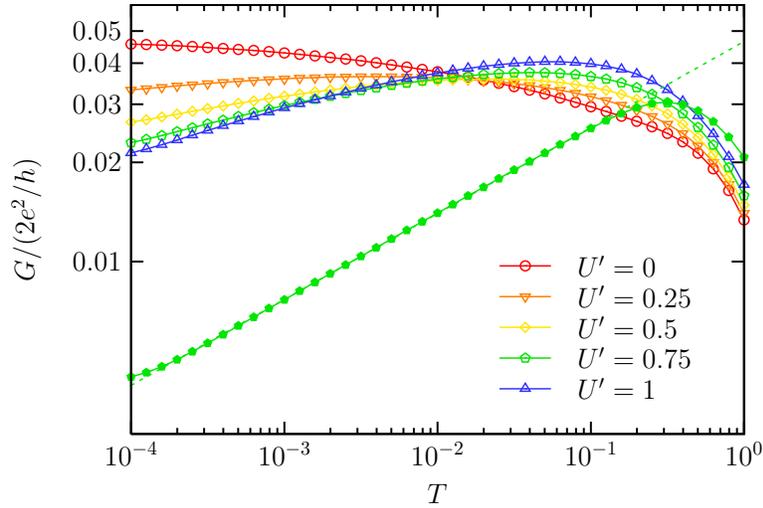}}
\vskip 5mm 
\caption{\label{fig:tg_extended} (Color online)
 Temperature dependence of the conductance for the extended Hubbard
 model with $L = 10^4$ sites and a single site impurity of strength 
 $V = 10$, for a Hubbard interaction $U = 1$ and various choices 
 of $U'$; 
 the density is $n=1/2$, except for the lowest curve,
 which has been obtained for $n=3/4$ and $U' = 0.65$ (leading to a 
 very small backscattering interaction);
 the dashed line is a power-law fit for the latter parameter set.}
\end{figure}

Fig.~\ref{fig:tg_extended} shows typical fRG results for the 
temperature dependence of the conductance for the extended Hubbard 
model with a single strong site impurity ($V=10$). 
Similar results were obtained for a hopping impurity.
The considered size $L=10^4$ corresponds to interacting wires in 
the micrometer range, which is the typical size of quantum wires 
available for transport experiments. 
For $U'=0$ the conductance \emph{increases} as a function of 
decreasing $T$ down to the lowest temperatures in the plot. 
For increasing nearest-neighbor interactions $U'$ a suppression
of $G(T)$ at low $T$ becomes visible, but in all the data obtained
at quarter-filling the suppression is much less pronounced than 
what one expects from the asymptotic power law with exponent 
$2\alpha_B$. 
By contrast, the suppression is much stronger and follows the
expected power law more closely if parameters are chosen such
that two-particle backscattering becomes negligible at low $T$,
as can be seen from the conductance curve for $n=3/4$ and 
$U'=0.65$ in Fig.~\ref{fig:tg_extended}. The value of $K_{\rho}$
for these parameters almost coincides with the one for another
parameter set in the plot, $n=1/2$ and $U'=0.75$, but the
behavior of $G(T)$ is completely different.
Note that at $T \sim \pi v_F/L$ finite size effects set in, 
as can be seen at the low $T$ end of some of the curves in the
figure.
An enhancement of the conductance due to backscattering has
been found already earlier in a renormalization group study
of impurity scattering in the g-ology model.\cite{YGM}

\begin{figure}[ht!]
\center{\includegraphics[width=10.cm]{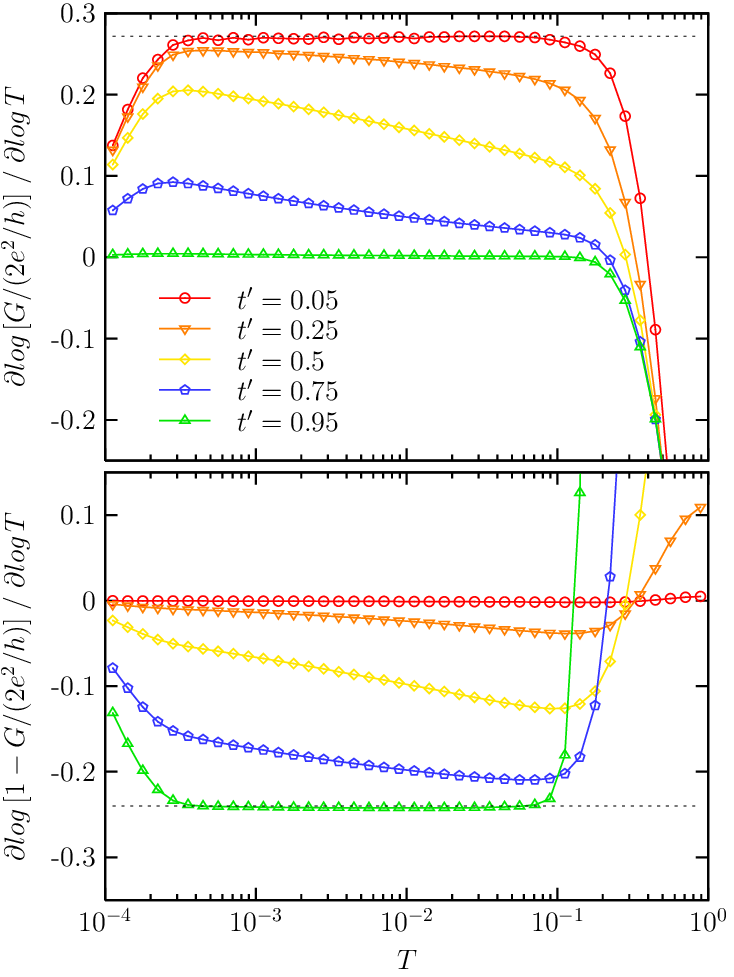}}
\vskip 5mm
\caption{\label{fig:condsingle} (Color online)
 Logarithmic temperature derivative of the conductance (upper panel)
 and of its deviation from the unitarity limit (lower panel) for the
 extended Hubbard model with $L = 10^4$ sites and various hopping
 impurities. The density is $n=3/4$, interaction parameters are
 $U=1$ and $U'=0.65$. The dashed horizontal lines highlight
 power-law behavior.}
\end{figure}

Results for the conductance of the extended Hubbard model with 
a hopping impurity with various amplitudes $t'$ are shown in 
Fig.~\ref{fig:condsingle}. The bulk parameters have been chosen 
such that the two-particle backscattering is practically 
zero at low $T$. From the plot of the logarithmic derivative 
of $G(T)$ in the upper panel one can see that for a strong
impurity (small $t'$) the conductance 
follows a well defined power law $G(T) \propto T^{2\alf_B}$
over a large temperature range. For intermediate $t'$ the
curves approach the asymptotic exponent at low $T$ from below, 
but do not reach it before finite size effects lead to a 
saturation of $G(T)$ for $T < \pi v_F/L$.
For the weakest impurity in the plot, $t'=0.95$, the conductance
remains very close to the unitarity limit. However, the plot
of the logarithmic derivative of $1 - G/(2e^2/h)$ in the lower
panel of Fig.~\ref{fig:condsingle} shows that $1 - G/(2e^2/h)$
increases as $T^{K_{\rho} - 1}$ for decreasing $T$, as expected
for a weak impurity in the perturbative regime.\cite{KF}
The effective exponents indicated by the two horizontal lines 
in the figure deviate from the exact values (determined from
the DMRG result \cite{EGN} for $K_{\rho}$) by about $20 \, \%$ 
in the case of $2\alf_B$ and only by $5 \, \%$ for $K_{\rho} - 1$.
Results for the conductance of a wire with a double-barrier
impurity at or near resonance will be presented in a forthcoming
publication.\cite{AEM}

\section{Conclusion}

We have derived a fRG-based computation scheme for the 
one-dimensional extended Hubbard model with a single static 
impurity, extending previous work for spinless fermions 
\cite{AEX,EMX} to spin-$\frac{1}{2}$ fermions. 
The underlying approximations are devised for weak short-range 
interactions and arbitrary impurity potentials.
Various observables have been computed: the local density of
states near boundaries and impurities, the density profile,
and the temperature dependence of the linear conductance.
Results have been checked against DMRG data,
for those observables and system sizes for which such data could
be obtained. The general agreement is good at weak coupling,
but for intermediate interaction strengths with sizable 
two-particle backscattering and strong impurities the deviations 
are significantly larger than for spinless fermions.
We suspect that the neglected influence of impurities on vertex
renormalization at high energy and short length scales is more 
important for fermions with spin.

Two-particle backscattering of particles with opposite spin at
opposite Fermi points leads to two important effects, not
present in the case of spinless fermions. First, the expected
decrease of spectral weight and of the conductance at low
energy scales is often preceded by an \emph{increase}, which 
can be particularly pronounced for the density of states near 
an impurity or boundary as a function of $\om$.
For the density of states near a boundary this effect has been
found already earlier within a Hartree-Fock and DMRG study of 
the Hubbard model,\cite{SMX,MMX} and for the conductance by a 
renormalization group analysis of the g-ology model.\cite{YGM}
Second, the asymptotic low-energy power laws are usually 
modified by \emph{logarithmic corrections}.
In the extended Hubbard model the backscattering can be 
eliminated for a special fine-tuned choice of parameters.
Then the results are very similar to those for spinless fermions.
For weak and intermediate impurity strengths the asymptotic
low-energy behavior is approached only at rather low scales,
which are accessible only for very large systems.
This slow convergence was observed already for spinless fermions
\cite{MMSS,AEX,EMX}
and holds also in the absence of two-particle backscattering.

For systems with long-range interactions backscattering is 
strongly reduced compared to forward scattering. 
This seems to be the case in carbon nanotubes.\cite{cnt} 
Hence, the conductance can be expected to follow the asymptotic 
power law at accessible temperature scales for sufficiently 
strong impurities in these systems, as is indicated also by 
experiments.\cite{Yao}
However, the effects due to two-particle backscattering should be 
observable in systems with a screened Coulomb interaction.

\vskip 1cm

\noindent
{\bf Acknowledgments:} \\
We thank Manfred Salmhofer for valuable discussions, Satoshi 
Nishimoto for providing the DMRG data of $K_{\rho}$ in the extended 
Hubbard model, and Roland Gersch for a critical reading of the
manuscript.
V.M.\ and K.S.\ are grateful to the Deutsche Forschungsgemeinschaft
(SFB 602) for financial support.

\vskip 1cm

\begin{appendix}

\section{Frequency cutoff at finite temperature}

In this appendix we derive a convenient implementation of a sharp frequency 
cutoff at finite temperature.  
The idea is roughly to rewrite the Matsubara sum as an integral over a 
piecewise constant function and then to introduce a sharp cutoff on this
continuous frequency as usual.

In the 1PI scheme \cite{SH} the flow equation with frequency cutoff at
zero temperature has the form (assuming no frequency shift along the
loop)\cite{Ens}
\begin{equation}
  \label{eq:cutoff:zeroT1}
  \frac{\partial\Upsilon^\Lambda}{\partial\Lambda}
  = \int \frac{d\omega}{2\pi} \;
  \delta_\epsilon(|\omega| - \Lambda) \;
  f\bigl(\Theta_\epsilon(|\omega| - \Lambda),\omega,\Upsilon^\Lambda\bigr)
  \; .
\end{equation}
Here $\Upsilon^\Lambda$ is the generating functional for the 1PI
vertex functions, and
$f(t,\omega,\Upsilon^\Lambda)$ represents the right-hand side of the flow 
equations including the momentum integral, but with the integral over the
Matsubara frequency $\omega$ written explicitly.  $\Theta_\epsilon(x)$
is a step function smoothed on a scale $\epsilon$, and
$\delta_\epsilon(x) = \partial_x \Theta_\epsilon(x)$.  The $\omega$
integral gives finite contributions near $|\omega| = \Lambda$, hence
$\Upsilon^\Lambda$ is a continuous function of $\Lambda$.  Then the
limit $\epsilon\to 0$ can be performed using Morris' lemma,\cite{Mor}
\begin{align}
  \label{eq:cutoff:zeroT}
  \frac{\partial\Upsilon^\Lambda}{\partial\Lambda}
  \overset{\epsilon\to 0}{\longrightarrow}
  \int \frac{d\omega}{2\pi} \;
  \delta(|\omega| - \Lambda)
  \int_0^1 dt \; f\bigl(t,\omega,\Upsilon^\Lambda\bigr)
  = \frac{1}{2\pi} \sum_{\omega=\pm\Lambda}
  \int_0^1 dt \; f\bigl(t,\omega,\Upsilon^\Lambda\bigr).
\end{align}
The right-hand side is independent of the explicit cutoff function
$\Theta_\epsilon(x)$, therefore the vertex functions in
$\Upsilon^\Lambda$ are smooth not only in $\Lambda$ but also as
functions of all external frequencies.

In our previous work at finite temperature\cite{EMX} the flow equation
\eqref{eq:cutoff:zeroT1} with discrete Matsubara frequencies reads
\begin{align}
  \label{eq:cutoff:finiteT}
  \frac{\partial\Upsilon^\Lambda}{\partial\Lambda} & = T \sum_n
  \delta_\epsilon(|\omega_n| - \Lambda) \;
  f\bigl(\Theta_\epsilon(|\omega_n| - \Lambda),\omega_n,
    \Upsilon^\Lambda\bigr).
\end{align}
In the limit $\epsilon\to 0$ the right-hand side contains a $\delta$ 
function and $\Upsilon^\Lambda$ jumps as $\Lambda$ passes $\omega_n$, hence 
the Morris lemma which requires continuity of $\Upsilon^\Lambda$ cannot be
applied.  We have therefore used a smooth cutoff, which renders the
numerics relatively slow.  Instead, we now propose a new
sharp cutoff which allows to apply the Morris lemma.  This reduces the
runtime significantly and enables us to access another order of magnitude 
in system size and temperature.

First, we rewrite the Matsubara sum as an integral over a continuous
frequency $\omega$ with a weight function peaked in non-overlapping 
neighborhoods of width $\eta$ around each $\omega_n$,
\begin{align}
  T \sum_n f(\omega_n)
  = \int d\omega \; T \sum_n \delta_\eta(\omega-\omega_n) \; f(M(\omega))
\end{align}
where $\int d\omega \, \delta_{\eta}(\omega - \omega_n) = 1$ and
$M(\omega)$ returns the discrete Matsubara frequency $\omega_n$
closest to $\omega$.  Hence, $f$ is a piecewise constant function of
a continuous variable $\omega$.  At this stage, the cutoff
function in $\omega$ is introduced in the bare action.  This leads to
the flow equation
\begin{align}
  \label{eq:cutoff:finiteTdouble}
  \frac{\partial\Upsilon^\Lambda}{\partial\Lambda} & = \int d\omega \;
  T \sum_n \delta_\eta(\omega-\omega_n) \; 
  \delta_\epsilon(|\omega| - \Lambda) \;
  f\bigl(\Theta_\epsilon(|\omega| - \Lambda),M(\omega),\Upsilon^\Lambda\bigr).
\end{align}
If we take the limit $\eta\to 0$ first we again obtain equation
\eqref{eq:cutoff:finiteT}, which is the smooth cutoff limit.  For
finite $\eta$, however, Morris' lemma gives
\begin{align}
  \frac{\partial\Upsilon^\Lambda}{\partial\Lambda}
  & \overset{\epsilon\to 0}{\longrightarrow}
  \int d\omega \;
  T \sum_n \delta_\eta(\omega-\omega_n) \; 
  \delta(|\omega| - \Lambda)
  \int_0^1 dt \; f\bigl(t,M(\omega),\Upsilon^\Lambda\bigr) \nonumber \\
  \label{eq:cutoff:finiteTeta}
  & = T \sum_n \delta_\eta(|\omega_n| - \Lambda)
  \int_0^1 dt \; f\bigl(t,\omega_n,\Upsilon^\Lambda\bigr).  
\end{align}
For each $\omega_n$ separately this is an autonomous differential
equation, hence the result is independent of the shape of
$\delta_\eta(x)$.  A convenient choice is a box of height $1/(2\pi T)$
and width $2\pi T$ centered around the Matsubara frequency $\omega_n$,
that is, $\delta_\eta(x)=1/(2\pi T)$ for $|x|<\pi T$ and $0$ otherwise,
which leads to the final result
\begin{align}
  \label{eq:cutoff:final}
  \frac{\partial\Upsilon^\Lambda}{\partial\Lambda}
  & = \frac{1}{2\pi} \sum_{\omega_n\approx\pm\Lambda}
  \int_0^1 dt \; f\bigl(t,\omega_n,\Upsilon^\Lambda\bigr).
\end{align}
Comparison with equation \eqref{eq:cutoff:zeroT} shows that the only
change necessary at finite temperature is to replace the loop
frequency $\omega=\pm \Lambda$ on the right-hand side by the nearest
discrete Matsubara frequency.  We have checked numerically that indeed
this new sharp cutoff gives the same results as the previous smooth
cutoff for the conductance $G(T)$ curves in a dramatically reduced
runtime.

\end{appendix}


\vfill\eject


\begin{thebibliography}{99}
\bibitem{dmrgspec} The DMRG data for the local spectral weight at 
 the Fermi level have been obtained by computing 
 $|\bra \psi^{(N-1)}|c_{j\up}|\psi^{(N)} \ket|^2$, where $\psi^{(N)}$ 
 is the $N$-electron ground state;
 the numerical error is below $5 \, \%$ in all shown results.
\end{thebibliography}
\end{document}